\documentclass[11pt,a4paper]{article}
\usepackage{jcappub}

\title{Fermi 130 GeV gamma-ray excess and dark matter annihilation in sub-haloes and in the Galactic centre}

\author[a,b]{Elmo Tempel,}
\author[a]{Andi Hektor}
\author[a,c,d]{and Martti Raidal}
\affiliation[a]{NICPB, Ravala 10, Tallinn 10143, Estonia}
\affiliation[b]{Tartu Observatory, Observatooriumi 1, T\~oravere 61602, Estonia}
\affiliation[c]{Institute of Physics, University of Tartu, Estonia}
\affiliation[c]{CERN, Theory Division, CH-1211 Geneva 23, Switzerland}
\emailAdd{elmo@aai.ee}
\emailAdd{andi.hektor@cern.ch}
\emailAdd{martti.raidal@cern.ch}

\abstract{We analyze publicly available Fermi-LAT high-energy gamma-ray data and confirm the existence of clear spectral feature peaked at $E_\gamma= 130$~GeV. Scanning over the Galaxy we identify several disconnected regions where the observed excess originates from. Our best optimized fit is obtained for the central region of Galaxy with a clear peak at 130~GeV with local statistical significance $4.5\sigma.$ The observed excess is not correlated with Fermi bubbles. We compute the photon spectra induced by dark matter annihilations into two and four standard model particles, the latter via two light intermediate states, and fit the spectra with data. Since our fits indicate sharper and higher signal peak than in the previous works, data favors dark matter direct two-body annihilation channels into photons or other channels giving only line-like spectra. If Einasto halo profile correctly predicts the central cusp of Galaxy, dark matter annihilation cross-section to two photons is of order ten percent of  the standard thermal freeze-out cross-section. The large dark matter two-body annihilation cross-section to photons may signal a new resonance that should be searched for at the CERN LHC experiments.}

\keywords{Gamma ray experiments, dark matter theory, particle physics -- cosmology connection}
\arxivnumber{1205.1045}

\notoc 
\begin{document}
\maketitle

\section{Introduction}

If the existing cosmological dark matter (DM)~\cite{Komatsu:11} is a thermal relic consisting of weakly interacting massive particles, DM annihilations into the standard model (SM) particles should provide indirect evidence of DM in cosmic ray experiments~\cite{Cirelli:11}. In this scenario the first emerging signal of DM annihilations is expected to appear from Galactic regions with the highest DM density such as the centre of Galaxy or the largest DM sub-haloes. Very recently it was claimed~\cite{Bringmann:12, Weniger:12} that there is a $4.6\sigma$ ($3.3\sigma$) local (global) evidence of a monochromatic gamma-ray line~\cite{Bergstrom:88, Bringmann:11} with an energy $E_\gamma\approx 130$~GeV present in the Fermi Large Area Telescope (LAT)~\cite{Atwood:09} publicly available data. The signal originates from the centre of Galaxy from a region obtained applying a signal-to-background optimization procedure on the gamma-ray data~\cite{Weniger:12}. If the claim is true, this could be the very first evidence 
that the DM is of particle physics origin, representing a breakthrough both in cosmology and in particle physics.

According to the analyses presented in ref.~\cite{Weniger:12}, fitting the Fermi-LAT excess with a narrow peak is more an assumption rather than a result. In fact, any sufficiently hard spectrum with sharp fall-off around 130~GeV, for example a box, would fit the data presented in figure~4 of~\cite{Weniger:12} as well as the narrow peak. This would open a possibility to explain the excess with astrophysical sources, for example with an unknown mechanism associated with the Fermi bubbles~\cite{Su:10}.  As the best signal-to-background regions determined in~\cite{Weniger:12} seem to overlap with the Fermi bubbles, such a qualitative connection is easy to come~\cite{Profumo:12}.  In addition, this would also allow to explain the observed excess with photon spectra from DM annihilations into standard model final states that produce significant amount of prompt photons in their decays~\cite{Ibarra:12}. Discriminating between those possibilities requires thorough study of the Fermi data.

In this work we analyze the publicly available Fermi-LAT gamma-ray data in order to check and study independently the claim of $E_\gamma\approx 130$~GeV excess. We analyze the Fermi-LAT 195 week \mbox{ULTRACLEAN} dataset using the kernel smoothing method for fitting that is independent of binning and is complementary to the sliding energy window method used by Weniger. To identify signal target regions we use data driven method similar to the one used by Fermi Collaboration in searches for DM sub-haloes~\cite{Ackermann:2012nb}.   We estimate the errors of our fits with the bootstrap method.  Using Monte Carlo method we study what is the probability that the observed excesses are statistical fluctuations of background. We are interested in finding out whether the excess exists, what is its spatial distribution in the Galaxy, what is its spectrum, possible origin etc.

Fitting the photon spectrum coming from the target regions identified by Weniger, we do confirm the existence a spectral feature in the analyzed data centered at $E_\gamma =130$~GeV. Knowing the energy of excess, we scan the data to find the regions where the excess comes from. We find that the excess originates from relatively small disconnected regions, the most important of them is the centre of Galaxy but several other regions exist. Away from the identified regions the excess disappears consistently with the expectations from DM annihilations. We fit the gamma-ray background at energies 20--300~GeV from data by cutting out the central signal region of the Galaxy. We obtain a perfect power-law fit with a power 2.6 for the high-energy gamma-ray background. We then fit the spectrum from the central signal region and observe a clear peak at $E_\gamma\approx 130$~GeV with a local statistical significance $4.5 \sigma$. Fits from other regions have local significances as high as  $3.2\sigma$. We comment on estimating the corresponding global statistical significances of those 
results.

Our results indicate that the shape of the peak is even more pronounced than obtained in~\cite{Weniger:12}: we observe also a rise of the peak. This result disfavors any possible explanation to the observed excess with shallow spectra (including the box-like spectra) over the power-law background, and favors more peaked   profiles that may be difficult to obtain from standard astrophysical sources. We study, classify and fit the possible model independent DM annihilation scenarios~\cite{Cirelli:09} into two and four standard model final states (the latter is assumed to occur via light intermediate states~\cite{ArkaniHamed:09} that may also induce Sommerfeld enhancement of the annihilation cross-section). We find that, among those spectra, our results disfavor any other scenario but direct two-body annihilation into $\gamma\gamma$ or $\gamma X$ final states, where $X$ is any massive particle. 
However, other narrow peak-like spectra like from internal bremsstrahlung~\cite{Bringmann:12} or 
narrow boxes due to massive intermediate particles~\cite{Ibarra:12}  are still allowed within present Fermi-LAT energy resolution.  
The DM annihilation cross-section to two photons should be of order ten percent of  the standard thermal freeze-out cross-section.
This result depends very sensitively on the unknown properties of the central cusp of our Galaxy and can therefore be relaxed. Increasing the loop-suppressed DM annihilation cross-section to photons may require the existence of new resonances that should be searched for at the CERN LHC experiments.

An important result of ref.~\cite{Weniger:12} is showing that regions with optimal size, depending on the assumed DM halo profile,  should be used for this type of search. Likely the reason why Fermi Collaboration did not observe the excess is that they looked at too large region~\cite{Abdo:10}. Assuming DM halo profiles introduces theoretical bias into the analyses.  In addition, we do not see a good reason why the low energy gamma-ray spectrum at energies $1~\mathrm{GeV} < E_\gamma <20$~GeV should be used to identify the best signal regions above $E_\gamma >100$~GeV. To the contrary, we believe that using the low energy spectrum for these purposes may be misleading. For example, we observe a slight asymmetry in the low energy spectrum that explains the north-south asymmetry of the identified regions in~\cite{Weniger:12}. Since for large regions the background is completely dominated by the Galactic disk, the requirement of good signal-to-background ratio cuts off the disk region. Thus the shape of the 
regions obtained in~\cite{Weniger:12} must trivially be of a hourglass type -- their overlap with the Fermi bubbles is most likely accidental. This coincidence lead some authors to speculate that the signal is associated with Fermi bubbles while the photons from the centre follow power-law background~\cite{Profumo:12}. Our results show that the actual situation with the $E_\gamma =130$~GeV excess is exactly opposite and the observed signal cannot be associated with the Fermi bubbles.

Based on that criticism we use data driven method similar to~\cite{Ackermann:2012nb} to search for signal regions, and we optimize their size ourselves. We find that the optimal signal region for the best statistical significance is covered by the radius of $3^\circ$ in the Galactic centre. For larger regions the signal significance decreases due to larger background, for smaller regions, like the Reg5 in~\cite{Weniger:12}, the signal decreases because of too small number signal photons from that small region. Thus we confirm the necessity of right choice of the signal regions. We also checked that the $3^\circ$ region in the centre of Galaxy where the significance of the signal over background is maximized is consistent with the expectations from DM annihilations for Einasto profile.  Our results show that the peak excess is concentrated to rather small regions. If the source of the 130~GeV photons is astrophysical, further investigation of those regions with different observation frequencies should reveal objects or processes that also produce the 130~GeV gamma-rays. However, if the origin of the excess is direct DM annihilation into photons, we may have identified the most dense DM regions of our Galaxy. This is anticipated result 
for the Galactic centre. However, for the other regions this claim must be confirmed by other experiments with more statistics because it is
possible that  we have observed just an upward statistical fluctuation of the background~\cite{Boyarsky:2012ca}.

\section{Data analyses}

\subsection{Data selection}

In the present analysis, we take into account 195 week (from 4 Aug 2008 to 18 April 2012) of data from the Fermi Large Area Telescope (LAT) with energies between 20 and 300~GeV.\footnote{http://fermi.gsfc.nasa.gov/ssc/data/access/} We apply the zenith-angle cut $\theta<100^\circ$ in order to avoid contamination with the earth albedo, as recommended by the Fermi LAT team. We also apply the recommended quality-filter cut \mbox{DATA\_QUAL}$=1$, \mbox{LAT\_CONFIG}$=1$, and \mbox{ABS(ROCK\_ANGLE)}$<52$.\footnote{http://fermi.gsfc.nasa.gov/ssc/data/analysis/LAT\_caveats.html} We make use of the \mbox{ULTRACLEAN} events selection (Pass~7 Version~6), in order to minimize potential statistical errors. The selection of events as well as the calculation of exposure maps is performed using  the 18 April 2012 version of ScienceTools \mbox{v9r27p1}.\footnote{http://fermi.gsfc.nasa.gov/ssc/data/analysis/} For everything else we use our own software.

In our analysis, we do not subtract the contribution from known point sources in the Fermi LAT data because the known astrophysical point sources are unlikely to be a problem at $E\ge 80$~GeV. However, we have checked that subtraction of known point-like sources do not affect our results in any way: the exclusion of known points sources do not affect the results when searching high-energy gamma-ray lines.

\subsection{Spectra estimation}
\label{sect:2.2}

Calculating the spectra from observed events is practically a probability distribution estimation. Estimating probability distributions is a well-developed topics in statistics, and we can choose all the modern tools -- kernel densities, adaptive kernels, smoothed bootstrap for point-wise confidence intervals. In~\cite{Tempel:11} this method is applied to study the luminosity function of galaxies. In this paper we shall use the same method, briefly reviewed below.

The simplest approach to find a observed spectra is the binned density histogram that depends both on the bin widths and the location of the bin edges. A better way to estimate a probability distribution is to use kernel smoothing (see, e.g.~\cite{Wand:95}), where the density is represented by a sum of kernels centered at the data points:
\begin{equation}
    \Phi(E)=\sum\limits_{i} \frac{1}{h_i}K\left( \frac{E-E_i}{h_i} \right).
\end{equation}
The kernels $K(x)$ are distributions $(K(x)>0, \int K(x)\mathrm{d}x=1)$ of zero mean and of a typical with $h$. The width is an analogue of the bin width, but there are no bin edges to worry about. In the latter equation, we use the adaptive kernel estimation, where the kernel widths depend on the data, $h_i=h(E_i)$. The summation is taken over all data points (events).

The kernel widths are known to depend on the density $f(x)$ itself, with $h\sim f(x)^{-0.2}$ (see, e.g.~\cite{Silverman:97}): in regions of fewer data points, we use wider kernels. This choice requests a pilot estimate for the density that can be found using a wide constant width kernel.

To estimate the spectra, we use $B_3$ box spline kernel:
\begin{equation}
B_3(x) = \frac{|x-2|^3 - 4|x-1|^3 + 6|x|^3 - 4|x+1|^3 + |x+2|^3}{12}.
\end{equation}
This kernel is well suited for estimating densities -- it is compact, differing from zero only in the interval $x\in [-2,2]$, and it conserves mass: $\sum_iB_3(x-i)=1$ for any $x$. This kernel is also close to Gaussian with $\sigma=0.6$.

We used the logarithmic energy scale for our spectra estimation. For the pilot estimate, we used a wide kernel with the scale $h=0.1$ (in logarithmic energy scale). This wide kernel leads very smooth distribution and is used to estimate the approximate event probability depending on energy. For the adaptive kernel widths, we adopted $h=0.03$ (the typical uncertainties in the overall energy calibration of the Fermi LAT) as the minimal width (for the maximum pilot density) and rescaled it by the $h\sim f_\mathrm{pilot}(x)^{-0.2}$ law. The event probability is higher in the low energies and drops at the high energies, leading to roughly twice as wide kernels for high energies in studied energy interval.


If we choose the kernel width this way, we minimize the mean integrated standard error (MISE) of the density. We are also interested in the ``error bars'', point-wise confidence limits (CL) for the density. This can be obtained by smoothed bootstrap~\cite{Davison:97, Fiorio:04}. Here the data points for the bootstrap realizations are chosen, as usual, randomly from the observed data with replacement, but they have an additional smoothing component:
\begin{equation}
    E_i^\star=E_j+h\epsilon_j,
\end{equation}
where $\epsilon$ is a random variable of the density $K(x)$.

We generated 10000 bootstrap realizations, using the adaptive kernel widths. We show the centered 95\% confidence regions in our figures.

Using the kernel smoothing method to estimate the probability distribution is appropriate when the potential signal is weak or the number of data points is small. In this case the usual binning technique may hide the signal or introduce a false signal, depending on the bin width and bin location. The kernel smoothing method depends only on the number of data points and the used kernel widths. Whereby, the kernel widths take into account the uncertainties of the observed data points and this way the resulting distribution function is effectively unaffected of it. As a result, the estimated probability distribution reveals the true observed distribution as accurately as possible.

\section{Fits to data: signal versus background}

Our aim is to perform as model independent fits to the Fermi LAT data as possible. To achieve this goal we first find the high-energy gamma-ray background from data. For that we exclude the central Galactic region with a radius of $12^\circ$ from data and fit all the data in the energy range $20~\mathrm{GeV}< E_\gamma <300$~GeV. The choice to exclude the central region of Galaxy from the background fit is motivated by the expected signal in this region. However, actually the result is insensitive to doing that. The fitting procedure is described in the previous section. We obtain almost perfect power-law background estimate from data with the power 2.6. This result agrees with qualitative theoretical expectations and shows that the gamma-ray background in this energy region is induced by scattering of high-energy diffuse protons. We use the data-fitted background in all our computations.

\begin{figure}
    \center
    \includegraphics[width=0.7\textwidth]{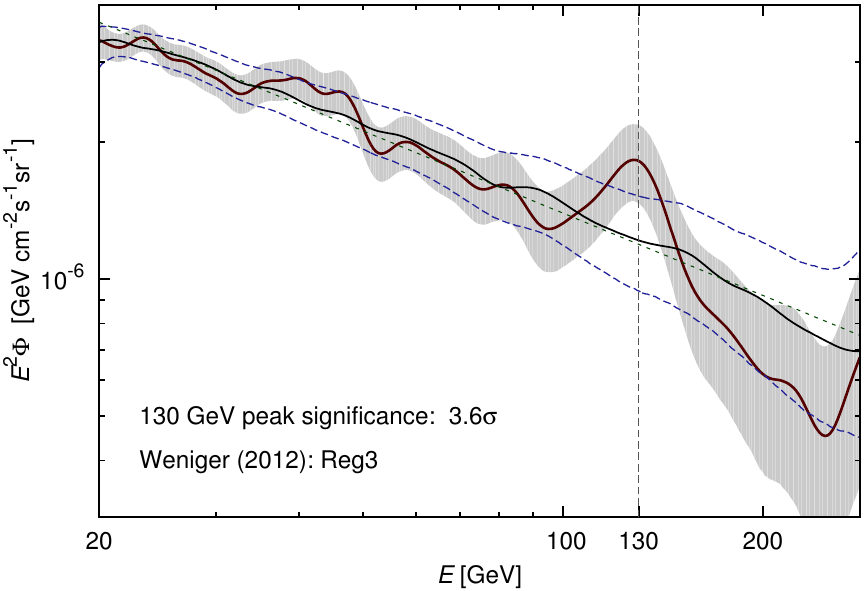}
    \caption{Estimated high-energy gamma-ray spectrum originating from Reg3 of ref.~\cite{Weniger:12}
     together with 95\% CL error band as a function of photon energy. Data fitted background (black solid line) is also shown, the pawer-law spectrum with power 2.6 (dotted line) is plotted for comparison. The blue dashed lines show 95\% CL for statistical fluctuations of the background.}
    \label{fig1}
\end{figure}

Logically the first step towards more thorough analyses is to check the claim of the existence of $E_\gamma=130$~GeV spectral feature in the data. For that we first choose the Reg3 identified in ref.~\cite{Weniger:12} as our signal region and fit the gamma-ray data from that region as described in the previous section. The result is shown in figure~\ref{fig1} where we plot the resulting gamma-ray spectrum (red solid line) as a function of photon energy for the high-energy region $20~\mathrm{GeV}< E_\gamma <300$~GeV. The 95\% CL error band is shown with a grey band around the best fit. The background obtained from data (black solid line) and the perfect power-law spectrum with power 2.6 (dotted line) are also shown. The total number of high-energy photons and the number of photons with energies $120~\mathrm{GeV}< E_\gamma <140$~GeV coming from this signal region are presented in table~\ref{table1}. The excess at $E_\gamma =130$~GeV with statistical significance\footnote{The statistical significance is 
estimated using the bootstrap realizations as described in section~\ref{sect:2.2}. The significance is given respect to the calculated background of the spectrum.} $3.6\sigma$ is clearly visible. We observe that the excess has a more pronounced peak-like shape than that presented in figure~4 of ref.~\cite{Weniger:12}. However, the overall flux we obtain is in good agreement with the gamma-ray flux obtained in ref.~\cite{Weniger:12}.

The question one may ask is whether the observed excess in figure~\ref{fig1} could entirely be due to statistical fluctuations of the background. Another way to estimate the significance of the peak is to generate background fluctuations with Monte Carlo for the number of photons in the target region. To do that we generated 10000 Monte Carlo realizations where the photon energy distribution follows the observed background: the high-energy background shape is assumed to be the same over the sky. For every realization we calculated the spectra as described in previous section. Based on 10000 Monte Carlo realizations we extracted 95\% confidence level limits for statistical fluctuations of the background.  The results  are presented in figure~\ref{fig1} with blue dashed lines. While all other small excesses and deficits around the background are consistent with expectations of fluctuations, the signal at 130~GeV clearly exceeds 95\% CL for background.

\begin{figure}
    \center
    \includegraphics[width=0.42\textwidth]{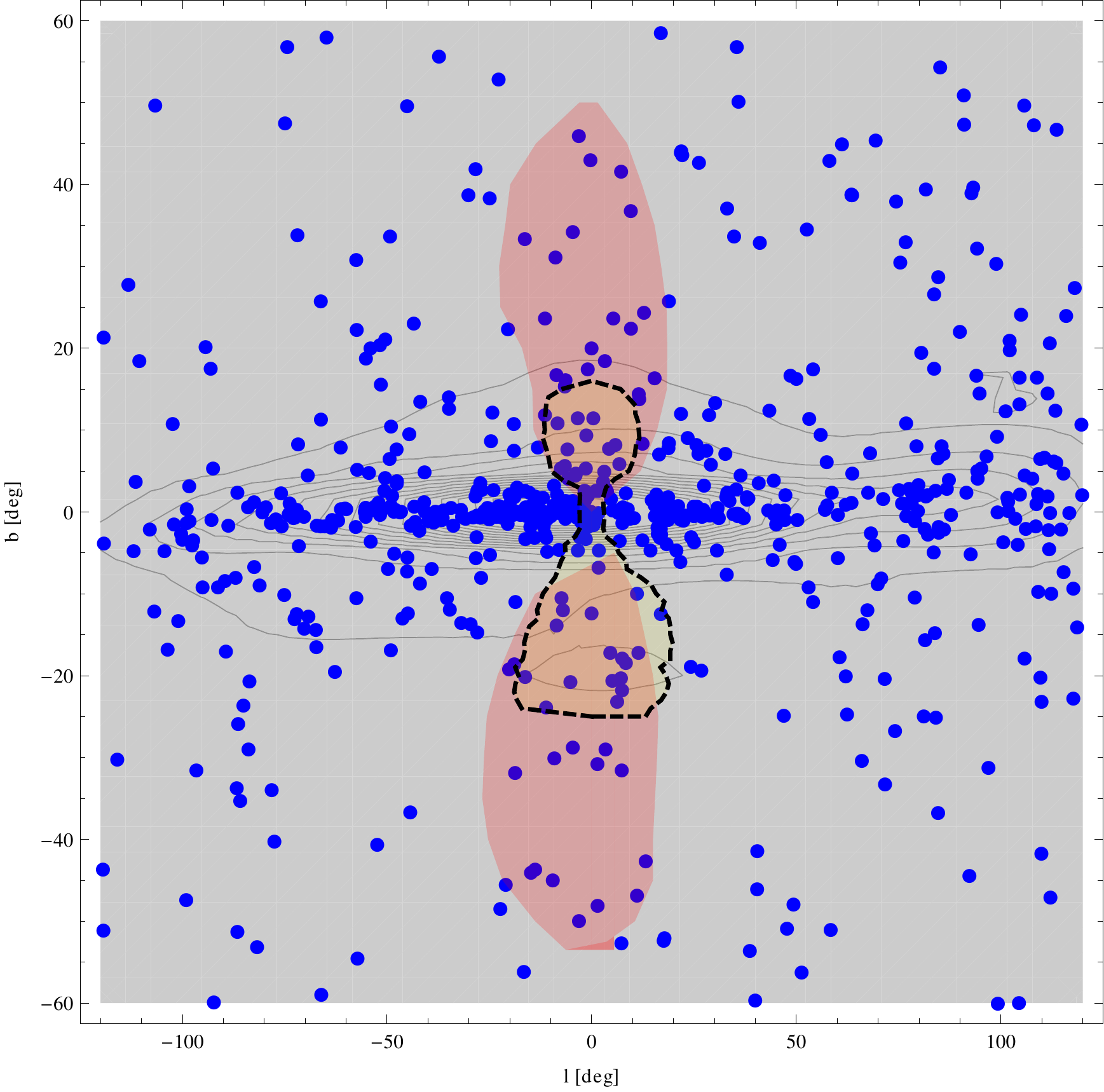}
    \includegraphics[width=0.57\textwidth]{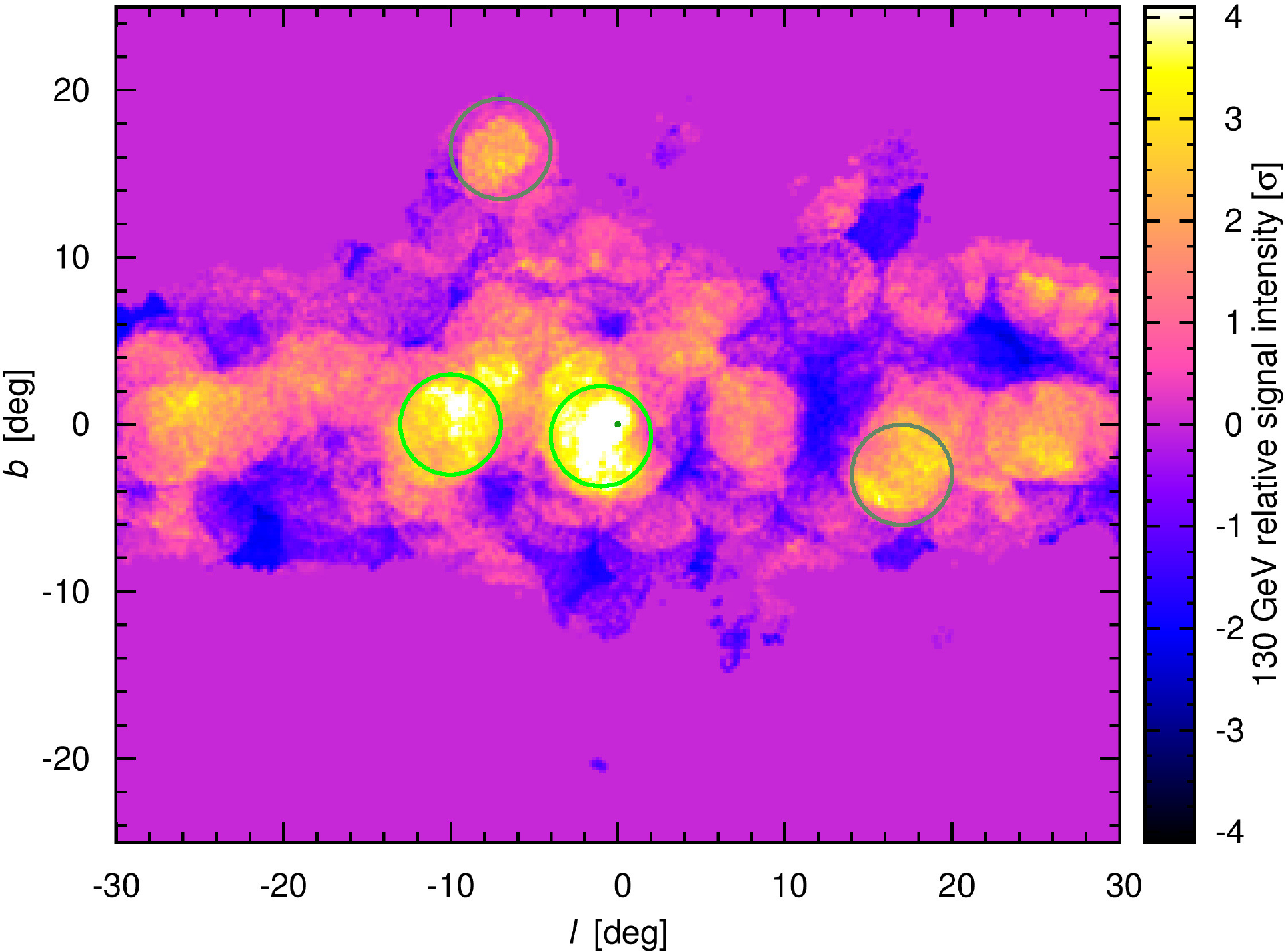}
    \caption{\emph{Left:} a Fermi ``photograph" of our Galaxy in gamma-rays with the energy $120~\mathrm{GeV}< E_\gamma <140$~GeV. Fermi data is shown with blue dots. The region Reg3 of ref.~\cite{Weniger:12} (dashed black curve) and the Fermi bubbles (red areas) are also shown for illustration. \emph{Right:} distribution of relative signal intensity of 130~GeV photons in the Galaxy. The green circles denote the signal regions that provide the excess with highest statistical significance; grey circles denote other regions showed in table~\ref{table1}; green dot mark the assumed centre of the Galaxy.}
    \label{fig2}
\end{figure}

Having confirmed the previous claims, the next question to ask is from which region of the Galaxy the photons in the peak come from? For illustration of data we first plot the distribution of photons in the energy range $120~\mathrm{GeV}< E_\gamma <140$~GeV, denoted with blue dots, in the left panel of figure~\ref{fig2}. Thus the figure is actually a Fermi photograph of our Galaxy in this energy range. As expected, we observe that most of the photons in that energy range come either form the very centre of the Galaxy or from the Galactic disk area.
  
In order to find the spatial origin of the 130~GeV excess we scan over the Galaxy choosing target regions with varying radii and fitting data coming from those regions as described before. Depending on the location we obtain either an excess or a deficit of the signal. At the same time we also compute 68\% CL for the background fluctuation for that region as just described: the background shape is assumed to be constant over the sky, and the CL depends only on the number of observed photons. We define relative signal intensity as a ratio of the signal excess/deficit and the 68\% confidence level background fluctuation limit (although it is in units of background fluctuation $\sigma$-s, we do not want to confuse the defined relative intensity with the statistical significances of the signal calculated with bootstrap). Scanning over the Galaxy we obtain a distribution of the relative signal intensity of the excess/deficit in the Galaxy. To eliminate very large statistical fluctuations, we calculate the 
signal intensity only for regions where the number of photons in energy range 20--300~GeV is larger than 80.

We plot in the right panel of figure~\ref{fig2} the resulting distribution of relative signal intensity as presented by the colour code. The pink background is due to regions with too low photon flux to obtain statistically meaningful results.  As seen in the figure, the signal with highest significance  originates from the centre of Galaxy. This region is centered at $(l,b)=(-1^\circ,-0.7^\circ)$, called ``Central" region in the following, and has a radius~$3^\circ,$ drawn with a white circle in figure~\ref{fig2}. The total number of high-energy photons and the number of $120~\mathrm{GeV}< E_\gamma <140$~GeV photons coming from this signal region is presented in table~\ref{table1}. However, there exist other regions, spatially well separated from the centre, that also exhibit large 130~GeV gamma-ray excess over the background. The most significant of them, with the same radius, is located at $(l,b)=(-10^\circ,0^\circ)$, called ``West" region in the following, and is also shown in the figure.  Some other 
possible  signal regions are all listed in table~\ref{table1}. Presently statistically significant fits are obtained only for the first two regions, but with more Fermi statistics the other regions may become relevant too.

One can see in figure~\ref{fig2} that the regions with excesses and the regions with deficit of the signal are not in balance -- the excess dominates (considering the degree of darkness/brightness of the regions).  The deficit almost never exceeds $2\sigma$ level and is in good agreement with the expectations from statistical fluctuations of the background. At the same time, there exist regions in which the observed excess exceeds the $2\sigma$ level.

It is clear from figure~\ref{fig2} that the excess of photons with energy around 130~GeV does not originate from Fermi bubbles. Firstly, there is no spatial correlation between the signal excess and the Fermi bubbles. Secondly, whatever is the physical mechanism creating the 130~GeV excess, this mechanism must be at work in several regions of the Galaxy. If the origin of the excess is astrophysical, it should be possible to observe those astrophysical objects/processes in the identified regions with other methods. Any such a mechanism must also explain why the observed excess is a peak, that might be difficult in the case of standard astrophysical processes. If, however, the origin of the 130~GeV peak is DM annihilations, figure~\ref{fig2} shows the distribution of the most dense DM sub-structures in the central region of our Galaxy. Notice that the dark centre of the Galaxy does not exactly coincide with the galactic coordinate origin.

\begin{figure}[t]
    \center
    \includegraphics[width=0.495\textwidth]{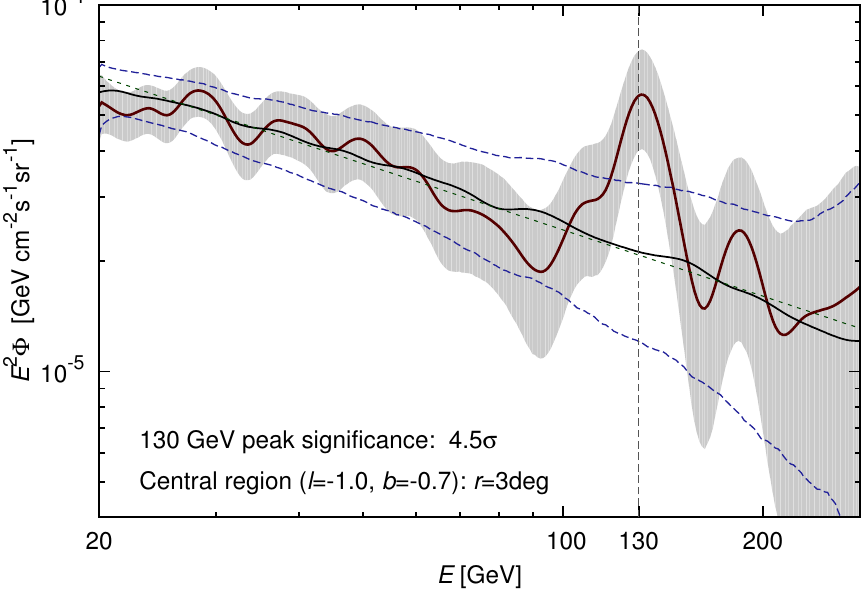}
    \includegraphics[width=0.495\textwidth]{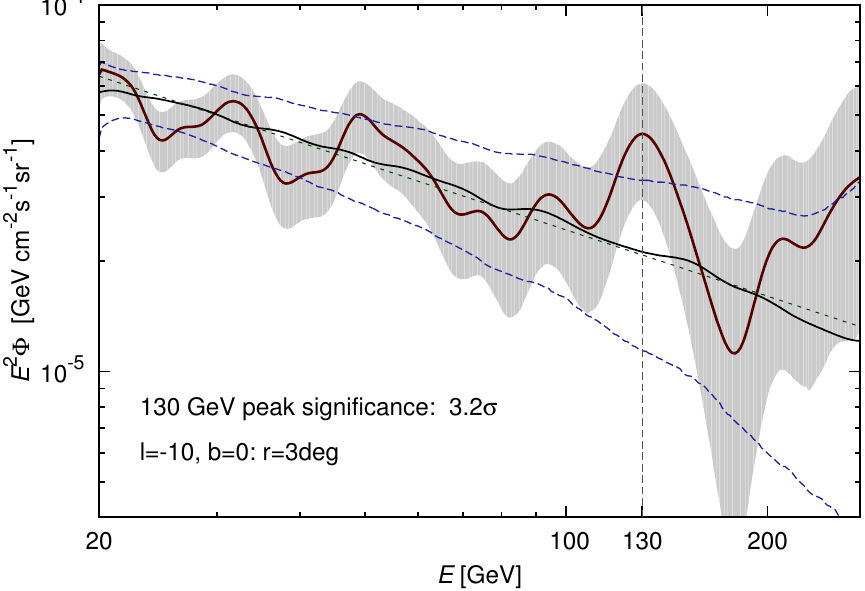}
    \caption{Best fits to high-energy gamma-ray data for the Central (\emph{left panel}) and West (\emph{right panel}) signal regions presented in table~\ref{table1}, together with 95\% CL error band as functions of photon energy. Background fitted from data is also shown (black solid line), the power-law spectrum with power 2.6 is plotted for comparison (dotted line). The blue dashed lines show 95\% CL for statistical fluctuations of the background.}
    \label{fig3}
\end{figure}

The fits to high-energy gamma-ray data originating from the Central and West signal regions are plotted in the left and right panels of figure~\ref{fig3}, respectively, using the same notation as in figure~\ref{fig1}. The Central region exhibits an excess with local statistical significance of $4.5\sigma$. To estimate the global significance we consider the trials factor $\simeq (\text{energy range under study})/(\text{width of the line})$, which reduces the global significance to $4.0\sigma$. The angular size of the region is placed by the null hypothesis of the annihilation signal from the DM halo profile. The difference of the signal in case of the Einasto and NFW profiles and also the off-axis location of the Central region can slightly reduce the global significance further. The fit to West region shows a clear peak at 130~GeV with local statistical significance of $3.2\sigma$. We have also fitted the signal from other bright regions in figure~\ref{fig2} that all show an excess peaked at the same photon energy, $E_\gamma=130$~GeV. Those are listed in table~\ref{table1}. The trial factor of the those regions is larger due to large region of sky scanned for the signal, allowing for the possibility that the excess in those regions is an upward fluctuation~\cite{Boyarsky:2012ca}. This issue must be clarified by future experiments.

Based on the model independent results presented in figures~\ref{fig1}--\ref{fig3} and in table~\ref{table1}, we conclude that, whatever is the physics origin of the excess, its significance is high, it has a clear peak shape, and it comes from the small region in the Galactic centre and possibly from 
several other small regions we have identified.

\begin{table}[t]
\caption{Identified signal regions in the Galaxy, number of photons in the two energy intervals and the statistical significance of excess in those regions. The radii of regions are all $3^\circ$ (except for Weniger Reg3).}
\begin{center}
\begin{tabular}{|c|c|c|c|c|c|}
\hline\hline
Region & $l$ (deg) & $b$ (deg) & $N_\gamma$  (20--300)~GeV & $N_\gamma$  (120--140~GeV) & significance \\
\hline\hline
Weniger Reg3 & -- & -- & 3298 &65 & $3.6\sigma$ \\
\hline 
Central & $-1$ &$-0.7$ & 818 & 27 & $4.5 \sigma$ \\
West & $-10$ & 0 & 726 & 21 & $3.2 \sigma$ \\
East & 17 & $-3$ &481 & 14 & $2.7 \sigma$ \\
North & $-7$ & 16.5 & 109 & 4 & $1.6 \sigma$ \\
\hline
\end{tabular}
\end{center}
\label{table1}
\end{table}

\section{Fitting DM annihilation spectra}

\subsection{Comparison of different annihilation channels}
\label{sect:4.1}

It is appealing to attempt to explain that the observed gamma-ray excess with DM annihilations in the Galactic centre and in the most dense DM sub-haloes around the centre of our Galaxy. Our approach is model independent as described in refs.~\cite{Cirelli:11, Cirelli:09}. We assume that the DM particles annihilate into two SM particles, $DM+DM\to SM+SM,$ where $SM=\gamma,\,e,\,\mu,\,\tau, q,\,W,\,Z,\,h,$ where $q$ denotes any light or heavy quark. The final state SM particles decay and/or radiate photons and light fermions from the final state radiation. In addition, we also allow DM annihilations into two light hypothetical finals states, $DM+DM\to V+V,$ that decay as $V\to\gamma\gamma\,,ee,\,\mu\mu.$ Those particles have been postulated to exist~\cite{ArkaniHamed:09} in order to explain the DM annihilations to lepton and not quark final states. In addition, those light particles may induce Sommerfeld enhancement of the annihilation cross-section. Both those features are needed to explain the PAMELA 
anomaly of positron flux with DM annihilation scenarios. Assuming those annihilation channels we have computed the resulting decay/hadronization chains and final state radiation with \mbox{PYTHIA}~\cite{Sjostrand:08}, and obtained the resulting prompt photon spectra. We use those prompt spectra to fit the observed gamma-ray excess. We neglect photons from inverse Compton scatterings between possible charged annihilation products and the Galactic and CMB photons as this spectrum is too distributed to explain the observed peaked excess.

To study which DM annihilation scenarios can explain the observed excess best we estimate the goodness of the fit as follows. To fit the peak we choose the signal energy range to be $50~\mathrm{GeV}< E_\gamma <200$~GeV and divide it into $n=100$ bins. We compute $\chi^2$ according to 
\begin{equation}
\chi^2= \sum_i^n \frac{(f^\mathrm{th}_i-f^\mathrm{obs}_i)^2}{\sigma_i^\mathrm{obs \,2}}, \label{chi2} 
\end{equation}
where $f^\mathrm{obs}_i $ and $\sigma^\mathrm{obs}_i $ are the observed flux and its error from our fits to data for a bin $i$, and $f^\mathrm{th}_i $ is a theory estimate of signal plus background computed for every annihilation channel. The theoretical spectrum is calculated the same way as observed spectrum (see section~\ref{sect:2.2}) to have comparable spectra. For every annihilation channel we find the minimal reduced $\chi^2/n$ that gives the best fit of that theoretical model to data.

\begin{figure}
    \center
    \includegraphics[width=0.495\textwidth]{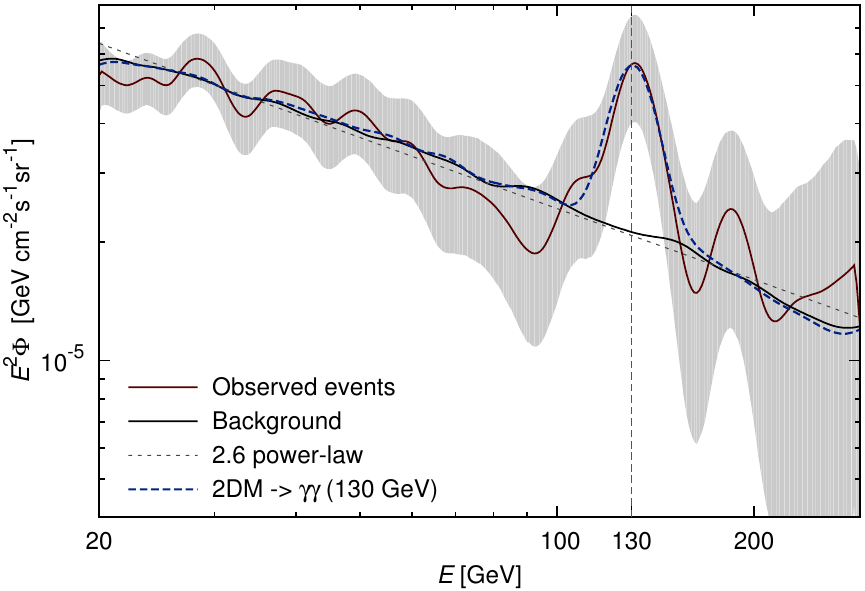}
    \includegraphics[width=0.495\textwidth]{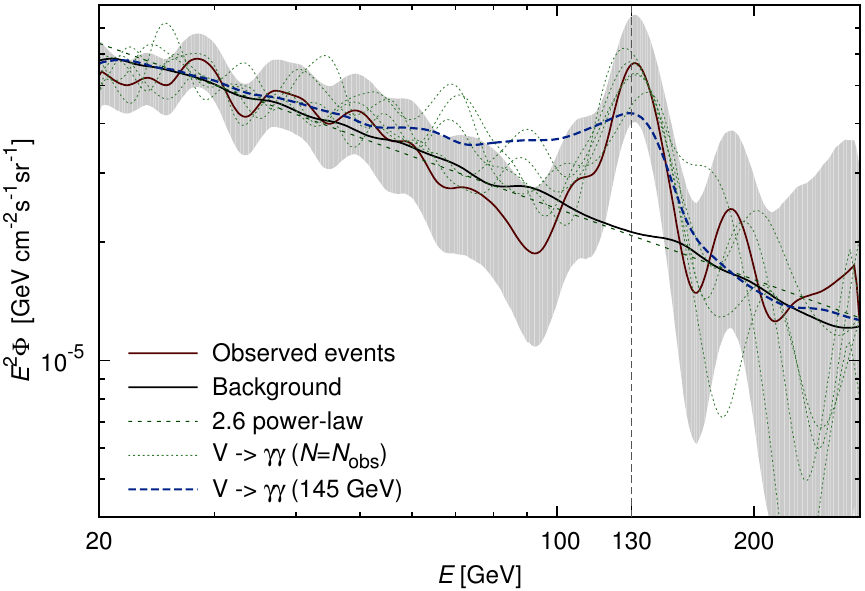}
    \caption{Best fit to data from the Central region with DM annihilation spectra into $2\gamma$ (\emph{left panel}) and to $2V\to4\gamma$ (\emph{right panel}), presented with dashed blue lines. The rest is as in figure~\ref{fig3}. The latter case can fit data only if large statistical fluctuation of the spectrum occur due to the present limited statistics, as demonstrated in the right panel with green dotted lines.}
    \label{fig4}
\end{figure}

Contrary to our initial intuition based on ref.~\cite{Weniger:12} results, almost all the resulting photon spectra from DM annihilations turn out to be too soft to explain the observed gamma-ray excess. Our best fit (for the Central signal region) for the annihilation channel $DM+DM\to \gamma+\gamma$, denoted by blue dotted line, is presented in the left panel of figure~\ref{fig4} together with the fit do data as in figure~\ref{fig3}. The best fit $\chi^2/n=0.7$ is obtained for the DM mass $M_\mathrm{DM}=130$~GeV. It was shown in ref.~\cite{Weniger:12} that DM annihilations into a monochromatic gamma-ray line can fit the data. 
Our results agree with this conclusion.

The second hardest photon spectrum is obtained for $DM+DM\to V+V\to 4\gamma$ channel, all the remaining studied channels produce so broad photon spectrum that cannot fit the observed peak. We present the best fit for this channel with $\chi^2/n=2.7$ in the right panel of figure~\ref{fig4} for the DM mass $M_\mathrm{DM}=145$~GeV. This annihilation channel can reproduce the fall-off of the peak but predicts flat box-like distribution for smaller than the peak energies. The reduced $\chi^2/n$ is 2.7 compared with 0.7 for the $2\gamma$ channel and the fit is clearly worse. It is still possible that, due to limited statistics, we observe significant down-ward fluctuation of the flux at photon energies $\sim 100$~GeV. With the present statistics the probability this to happen is at the level of a few percent. Examples of such fluctuations are presented in the right panel of figure~\ref{fig4} with green dotted lines. If we observe the down-ward fluctuation, this should disappear with more data and the observed peak 
will be replaced by a box-like spectrum. This scenario looks, however, unlikely.

In the previous computations we assumed almost massless intermediate particles $V$. However, the width of the box-like spectrum depends on the 
mass difference between the DM mass and the intermediate state mass~\cite{Ibarra:12}. For the fine tuned case $M_\mathrm{DM}  \sim M_V$ the box-like spectrum shrinks into a narrow peak-like spectrum, and a good fit to data becomes possible. 

\subsection{Fitting annihilation cross-section}

Up to now we have made no assumption about the DM distribution in the main DM halo of the Galaxy. However, if we want to estimate the annihilation cross-sections that fit the signal, we must compute the absolute flux of gamma-rays coming from the signal regions. Here we work only with the Central signal region. We assume two different halo profiles, the Einasto profile~\cite{Einasto:65,Navarro:04,Springel:08,Pieri:11},
\begin{equation}
 \rho_\mathrm{Ein}(r) = \rho_s \exp\left\{-\frac{2}{\alpha}\left[\left(\frac{r}{r_s}\right)^{\alpha}-1\right]\right\},
\end{equation}
with $\alpha=0.17$, $\rho_s=0.079$, and $r_s=20$ and the Navarro-Frenk-White (NFW) profile~\cite{Abdo:10, Navarro:97},
\begin{equation}
 \rho_\mathrm{NFW}(r) = \rho_s \frac{r_s}{r} \left(1+\frac{r}{r_s}\right)^{-2},
\end{equation}
with $\rho_s=0.33$ and $r_s=20$. The profiles have been normalized to the local DM density $\rho_\mathrm{DM} = 0.4$~GeV~cm$^{-3}$ at the Solar system~\cite{Catena:10,Salucci:10}. The gamma-ray flux from Central signal region is calculated according to formalism presented in~\cite{Cirelli:11}. To estimate the annihilation cross-section, $\langle \sigma \upsilon \rangle$, we let it to vary and compute the corresponding $\chi^2/n$ of the flux according to~\eqref{chi2}. The minimal $\chi^2/n$ gives the best fit value for $\langle \sigma \upsilon \rangle$.
To estimate the confidence limits for cross-section, we use the bootstrap sample calculated from the observed events. To every bootstrap realization, we fit the model spectra as described in section~\ref{sect:4.1} and find the best-fitted cross-section, $\langle \sigma \upsilon \rangle$. All bootstrap realizations give the distribution for $\langle \sigma \upsilon \rangle$, where we extract the 95\% confidence level limits given in table~\ref{tab2}.

The results for $2 DM \to 2 \gamma$ and $2 DM \to 2 V \to 4 \gamma$ annihilation channels with very light $V$  are presented in table~\ref{tab2}. They correspond to the cases plotted in figure~\ref{fig4}. 
The annihilation cross-sections to photons are required to be large, of order ten percent of standard thermal cross-section, with quite large errors. 
Within errors our result for annihilation cross section agrees with the result obtained by Weniger.

The resulting DM annihilation cross-section is much larger than expected in most of DM models. The result can be explained by cross-section enhancement mechanisms like the Sommerfeld enhancement or resonances in the annihilation process. We will elaborate on the latter in the next section. The enhancement of the flux can also be explained by non-standard DM cusp in the centre of Galaxy. In this case the annihilation cross-section can be small with the price of making the central cusp DM density bigger that predicted by the profiles we assume. If the main halo has substructures in the central region as favored by our results the presented cross-sections can have larger uncertainties as presented here. The over-dense substructures can explain the enhancement of 
the observed flux.  In this case, assuming some theoretical value 
$\langle \sigma \upsilon \rangle$ for the annihilation cross-section, our results  allow to compute the required DM density in the cusp of Galaxy
and in the most dense DM sub-haloes that induces  the measured flux.

\begin{table}[t]
\caption{The annihilation cross-sections for Central region given in units of the standard thermal cross-section, 
$\langle \sigma \upsilon \rangle = 3 \times 10^{-26}$ cm$^3$ s$^{-1}$ together with 95\% confidence level limits.}
\begin{center}
\begin{tabular}{|c|c|c|}
\hline\hline
Channel & $\langle \sigma \upsilon \rangle$ for Einasto & $\langle \sigma \upsilon \rangle$ for NFW \\
\hline\hline
$2 DM \to 2 \gamma$ & $0.09\pm{0.05}$ & $0.17\pm{0.09}$ \\
$2 DM \to 2 V \to 4 \gamma$ & $0.13\pm{0.09}$ & $0.25\pm{0.17}$ \\
\hline
\end{tabular}
\end{center}
\label{tab2}
\end{table}

\section{Discussion of the results}

While we do confirm the existence of the 130~GeV excess in Fermi LAT data, our results differ from the ones presented by ref.~\cite{Weniger:12} in some important aspects. Firstly, our search strategy for finding the most sensitive signal regions in the Galaxy gives significantly different results than obtained by Weniger. We do not  use the low energy gamma-ray data, $1~\mathrm{GeV}< E_\gamma <20$~GeV, to determine the high-energy gamma-ray target regions. At such low energies the background may be distorted by astrophysical sources like Fermi bubbles and, indeed, we observe north-south asymmetry in the low energy data. This explains why the Weniger target regions are asymmetric. In addition, Weniger's signal regions similarity with Fermi bubbles is most likely accidental but misleading. At the end, the fit to data from those regions gives quite significant excess, see figure~\ref{fig1}. Instead, we used a data driven method  to find the best target regions and we fitted the background from high-energy data 
by excluding the centre of Galaxy.  With this procedure we identified small regions where the signal significantly exceeds the background, presented in figure~\ref{fig2} and table~\ref{table1}. Similar method was
used by Fermi Collaboration to search for DM sub-haloes. In the context of our approach, the chosen target regions 
may correspond to DM sub-haloes of our Galaxy.

Secondly, our results show that the excess has clearly a shape of the peak while Weniger's data analyses may also be consistent with a box-like excess. This result narrows the possible astrophysical explanations to the excess. This result also disfavors all DM annihilation modes but the ones to photons
with narrow peak-like spectrum  (including internal bremsstrahlung~\cite{Beacom:05, Bergstrom:05, Bergstrom:05a} spectrum and narrow box-like
spectrum). At the moment the width of the allowed peak is mostly determined by the Fermi LAT energy resolution and quite poor statistics. 

We found that the  DM annihilation cross-section $\langle \sigma v \rangle$ should be  larger than naively expected for loop-induced annihilation processes. 
Indeed, typical suppression for  one loop processes involving weakly interacting particles is of order $10^{-4}$ compared to the typical tree level processes.  There are two possible ways to explain this result. Firstly, the DM annihilation cross-section into $\gamma\gamma$ may be small, but the DM halo models we have used, Einasto and NFW, fail to describe the central cusp of our Galaxy. This is possible since N-body simulations cannot predict the central region precisely. Since the cross-section calculation depends on the halo properties, there is always large theoretical uncertainty related to that. Secondly, the annihilation cross-section into photons might be, indeed, enhanced. Thus the model building of DM theories should concentrate on $\gamma\gamma$ or $\gamma X$ annihilation modes and on the enhancement of those annihilation cross-sections to the observed level.

Generically one expects DM direct annihilation cross-section to photons to be orders of magnitude smaller than is needed to explain the observation. However, examples exist in which loop level~\cite{Ferrer:2006hy,Gustafsson:07, Goodman:11, Profumo:10, Jackson:10,Chalons:2011ia} or anomaly induced~\cite{Dudas:09, Mambrini:09} cross-sections to photons are enhanced. Motivated by the recent LHC results that at least one light fundamental scalar, the Higgs boson, likely exist with a mass 125~GeV, and no low scale supersymmetry nor Kaluza-Klein particles exist below 1~TeV, one can construct most plausible scenarios of thermal relic DM that have large annihilation cross-sections to $2\gamma$ channel.

For example, one well motivated possibility is that the DM consists of dark scalars that couple to the SM fermion via extended Higgs sector with non-vanishing (or dominant) coupling only to the top quark. In this case the correct DM thermal relic abundance is induced via s-channel annihilation process for DM masses somewhat below the top quark mass~\cite{Jackson:10}. This is a generic prediction of the set-up not related to any model building detail but to the topology of freeze-out process and to the top mass. Today, when the DM is essentially at rest, the only kinematically allowed annihilation channel to the SM particles is opened by the top loop induced coupling to two photons (that dominates over the $\gamma Z,$ $ZZ$ final states). The kinematically allowed annihilation channels via virtual top-quarks imply 6-particle final states and are negligible.
This, as argued before, must be enhanced. The enhancement can occur due to an accidental resonance so that $2 M_\mathrm{DM}\sim M_S,$ where $M_S$ is the mass of new unobserved scalar. In this case all the parameters of the scenario are essentially fixed. The DM mass 130~GeV is consistent with 
thermal freeze-out, and new particle with non-standard couplings is predicted to exist with a mass $M_S\sim 260$~GeV, depending somewhat on how close to the resonance the annihilation process must be. A drawback of the scenario is that the DM-nucleon scattering cross-section is predicted to be small, explaining the absence of signal in DM direct searches at \mbox{XENON100}~\cite{Aprile:11}.

This scenario has already been realized in ref.~\cite{Jackson:10} for the $\gamma h$ final state, where $h$ is the SM Higgs boson, in the context of Randall-Sundrum framework with extra $Z'$ boson. For the final states $\gamma X$ the 130~GeV gamma-ray excess implies different DM mass that depends on the mass of $X$ via $E_\gamma=M_\mathrm{DM} (1- M_X^2/4M_\mathrm{DM}^2)$. Apart from the mass shift, to enhance the gamma-ray signal to the indicated level, all this type scenarios should include new resonance at $2M_\mathrm{DM}$ that can be searched for at the LHC experiments.

\section{Conclusions}

We have analyzed Fermi LAT publicly available data collected during 195 weeks and confirm the existence of gamma-ray excess peaked at $E_\gamma=130$~GeV. The excess originates from disconnected spatial regions, presented in table~\ref{table1}. The excess is not correlated with Fermi bubbles. The strongest signal over background comes from the central region of the Galaxy and has local statistical significance of $4.5\sigma$ (global $4.0\sigma$ after trials factor correction for energy), but the signal from the other regions shows also up to $3\sigma$ excess (in local significance). According to our fits the excess is narrower and higher than shown before. Leaving aside the possibility that the excess is an instrumental artefact, our results show that the mechanism of generating such an excess must be at work in several regions of our Galaxy. It might be difficult to explain the sharp gamma-ray peak with standard astrophysical processes. It is more appealing to assume that the excess originates from DM annihilations. In this case our results imply that we have identified the most dense DM substructures of our Galaxy -- the central cusp and possibly some DM sub-haloes, although the latter claim must be confirmed with better statistics in future experiments.

Assuming the DM annihilation scenario, we have computed photon spectra from DM annihilations into two SM particles and into two light bosons $V$ that decay to two photons or leptons using PYTHIA. We find that only the DM annihilations into $\gamma\gamma$ or $\gamma X$ final states, where $X$ is any massive particle,  can fit data well. All other spectra, including the $4\gamma$ one from light intermediate $V$ states, provide significantly worse fits to data. The exceptions are the possibilities that the $V$  mass is almost degenerate with the DM particle mass since then the box-like spectrum shrinks into a peak, and the
internal bremsstrahlung like spectra. We obtain that the DM annihilation cross-section into photons must be of order ten percent of the standard thermal freeze-out cross-section, see table~\ref{tab2}. However, this result is correct only if the Einasto or NFW halo profiles predict the properties of the central cusp exactly. Conversely, assuming some theoretically motivated DM annihilation cross-section to photons, our results allow to calculate the DM density in the cusp of Galaxy and in the most dense DM sub-haloes. We have sketched a generic thermal relic DM scenario that, independently of model details, should produce the observed DM relic density and the enhanced DM annihilation cross-section  into $\gamma\gamma$ today. This scenario predicts a new resonance close to $2 M_\mathrm{DM}$ with specific couplings that should be searched for at the LHC experiments.

\acknowledgments

We thank M.~Cirelli, E.~Gabrielli, C.~Grojean, T.~Hambye, G.~Servant, A.~Strumia and C.~Weniger, for discussions and communications. We thank J.~Pelt for constructive comments how to estimate the errors for cross-sections and J.~Cline and M.~McCullough  for pointing out a mistake in the photon spectrum in the first version of the manuscript. This work was supported by the ESF grants 8090, 8499, 8943, MTT8, MTT59, MTT60, MJD52, MJD272, by the recurrent financing projects SF0690030s09, SF0060067s08 and by the European Union through the European Regional Development Fund.


\newpage

\begin{thebibliography}{10}

\bibitem{Komatsu:11}
E.~{Komatsu}, K.~M. {Smith}, J.~{Dunkley}, C.~L. {Bennett}, B.~{Gold},
  G.~{Hinshaw}, N.~{Jarosik}, D.~{Larson}, M.~R. {Nolta}, L.~{Page}, and et al.,
  {\it {Seven-year Wilkinson Microwave Anisotropy Probe (WMAP)
  Observations: Cosmological Interpretation}},  {\em ApJS} {\bf 192} (2011) 18, [\href{http://xxx.lanl.gov/abs/1001.4538}{{\tt arXiv:1001.4538}}].

\bibitem{Cirelli:11}
For a review and for details of DM annihilation signals in cosmic ray spectra see,\\
M.~{Cirelli}, G.~{Corcella}, A.~{Hektor}, G.~{H{\"u}tsi}, M.~{Kadastik},
  P.~{Panci}, M.~{Raidal}, F.~{Sala}, and A.~{Strumia}, {\it {PPPC 4 DM ID: a
  poor particle physicist cookbook for dark matter indirect detection}},  {\em
  J. Cosmology Astropart. Phys.} {\bf 3} (2011) 51,
  [\href{http://xxx.lanl.gov/abs/1012.4515}{{\tt arXiv:1012.4515}}].

\bibitem{Bringmann:12}
T.~{Bringmann}, X.~{Huang}, A.~{Ibarra}, S.~{Vogl}, and C.~{Weniger}, {\it
  {Fermi LAT Search for Internal Bremsstrahlung Signatures from Dark Matter
  Annihilation}}, (2012),  [\href{http://xxx.lanl.gov/abs/1203.1312}{{\tt
  arXiv:1203.1312}}].

\bibitem{Weniger:12}
C.~{Weniger}, {\it {A Tentative Gamma-Ray Line from Dark Matter Annihilation at
  the Fermi Large Area Telescope}}, (2012),
  [\href{http://xxx.lanl.gov/abs/1204.2797}{{\tt arXiv:1204.2797}}].

\bibitem{Bergstrom:88}
L.~{Bergstr{\"o}m} and H.~{Snellman}, {\it {Observable monochromatic photons
  from cosmic photino annihilation}},  {\em Phys.~Rev.~D} {\bf 37} (1988)
  3737.

\bibitem{Bringmann:11}
For complete set of references see,\\
T.~{Bringmann}, F.~{Calore}, G.~{Vertongen}, and C.~{Weniger}, {\it {Relevance
  of sharp gamma-ray features for indirect dark matter searches}},  {\em
  Phys.~Rev.~D} {\bf 84} (2011) 103525,
  [\href{http://xxx.lanl.gov/abs/1106.1874}{{\tt arXiv:1106.1874}}].

\bibitem{Atwood:09}
W.~B. {Atwood}, A.~A. {Abdo}, M.~{Ackermann}, W.~{Althouse}, B.~{Anderson},
  M.~{Axelsson}, L.~{Baldini}, J.~{Ballet}, D.~L. {Band}, G.~{Barbiellini}, and
  et~al., {\it {The Large Area Telescope on the Fermi Gamma-Ray Space Telescope
  Mission}},  {\em ApJ} {\bf 697} (2009) 1071,
  [\href{http://xxx.lanl.gov/abs/0902.1089}{{\tt arXiv:0902.1089}}].

\bibitem{Su:10}
M.~{Su}, T.~R. {Slatyer}, and D.~P. {Finkbeiner}, {\it {Giant Gamma-ray Bubbles
  from Fermi-LAT: Active Galactic Nucleus Activity or Bipolar Galactic Wind?}},
   {\em ApJ} {\bf 724} (2010) 1044,
  [\href{http://xxx.lanl.gov/abs/1005.5480}{{\tt arXiv:1005.5480}}].

\bibitem{Profumo:12}
S.~{Profumo} and T.~{Linden}, {\it {Gamma-ray Lines in the Fermi Data: is it a
  Bubble?}}, (2012),  [\href{http://xxx.lanl.gov/abs/1204.6047}{{\tt arXiv:1204.6047}}].

\bibitem{Ibarra:12}
When our computations were finalized, this paper appeared that used similar idea to fit the gamma-ray excess with a box-like spectrum,\\
A.~{Ibarra}, S.~{L{\'o}pez Gehler}, and M.~{Pato}, {\it {Dark matter
  constraints from box-shaped gamma-ray features}}, (2012), 
  [\href{http://xxx.lanl.gov/abs/1205.0007}{{\tt arXiv:1205.0007}}].

\bibitem{Ackermann:2012nb}
  M.~Ackermann {\it et al.}  [The Fermi LAT Collaboration],
  ApJ  {\bf 747} (2012) 121,
  [\href{http://xxx.lanl.gov/abs/1201.2691}{{\tt arXiv:1201.2691}}].


\bibitem{Cirelli:09}
M.~{Cirelli}, M.~{Kadastik}, M.~{Raidal}, and A.~{Strumia}, {\it
  {Model-independent implications of the e, p$^{-}$ cosmic ray spectra on
  properties of Dark Matter}},  {\em Nuclear Physics B} {\bf 813} (2009)
  1, [\href{http://xxx.lanl.gov/abs/0809.2409}{{\tt arXiv:0809.2409}}].

\bibitem{ArkaniHamed:09}
N.~{Arkani-Hamed}, D.~P. {Finkbeiner}, T.~R. {Slatyer}, and N.~{Weiner}, {\it
  {A theory of dark matter}},  {\em Phys.~Rev.~D} {\bf 79} (2009) 015014,
  [\href{http://xxx.lanl.gov/abs/0810.0713}{{\tt arXiv:0810.0713}}].

\bibitem{Abdo:10}
A.~A. {Abdo}, M.~{Ackermann}, M.~{Ajello}, W.~B. {Atwood}, L.~{Baldini},
  J.~{Ballet}, G.~{Barbiellini}, D.~{Bastieri}, K.~{Bechtol}, R.~{Bellazzini},
  et~al., and {Fermi LAT
  Collaboration}, {\it {Fermi Large Area Telescope Search for Photon Lines from
  30 to 200 GeV and Dark Matter Implications}},  {\em Physical Review Letters}
  {\bf 104} (2010) 091302, [\href{http://xxx.lanl.gov/abs/1001.4836}{{\tt
  arXiv:1001.4836}}].
  

\bibitem{Boyarsky:2012ca}
  A.~Boyarsky, D.~Malyshev and O.~Ruchayskiy,
 {\it { Spectral and spatial variations of the diffuse gamma-ray background in the vicinity of the Galactic plane and possible nature of the feature at 130 GeV,}}
 [\href{http://arXiv.org/abs/arXiv:1205.4700}{{\tt arXiv:1205.4700}}].


\bibitem{Tempel:11}
E.~{Tempel}, E.~{Saar}, L.~J. {Liivam{\"a}gi}, A.~{Tamm}, J.~{Einasto},
  M.~{Einasto}, and V.~{M{\"u}ller}, {\it {Galaxy morphology, luminosity, and
  environment in the SDSS DR7}},  {\em Astron.~\&~Astrophys.} {\bf 529} (2011) A53, [\href{http://xxx.lanl.gov/abs/1012.1470}{{\tt arXiv:1012.1470}}].

\bibitem{Wand:95}
M.~P. Wand and M.~C. Jones, {\em Kernel Smoothing}.
\newblock Chapman \& Hall, (1995).

\bibitem{Silverman:97}
B.~W. Silverman, {\em Density Estimation for Statistics and Data Analysis}.
\newblock Chapman \& Hall, (1997).

\bibitem{Davison:97}
A.~C. Davison and D.~V. Hinkley, {\em Bootstrap Methods and their Application}.
\newblock Cambridge University Press, (1997).

\bibitem{Fiorio:04}
C.~V. Fiorio, {\it Confidence intervals for kernel density estimation},  {\em
  The Stata Journal} {\bf 4} (2004). 168.


\bibitem{Sjostrand:08}
T.~{Sj{\"o}strand}, S.~{Mrenna}, and P.~{Skands}, {\it {A brief introduction to
  PYTHIA 8.1}},  {\em Computer Physics Communications} {\bf 178} (2008)
  852, [\href{http://xxx.lanl.gov/abs/0710.3820}{{\tt arXiv:0710.3820}}].

\bibitem{Einasto:65}
J.~Einasto, {\it On the construction of a composite model for the galaxy and on
  the determination of the system of galactic parameters},  {\em Trudy Inst.
  Astrofiz. Alma-Ata} {\bf 5} (1965) 87.

\bibitem{Navarro:04}
J.~F. {Navarro}, E.~{Hayashi}, C.~{Power}, A.~R. {Jenkins}, C.~S. {Frenk},
  S.~D.~M. {White}, V.~{Springel}, J.~{Stadel}, and T.~R. {Quinn}, {\it {The
  inner structure of {$\Lambda$}CDM haloes - III. Universality and asymptotic
  slopes}},  {\em MNRAS} {\bf 349} (2004) 1039,
  [\href{http://xxx.lanl.gov/abs/astro-ph/0311231}{{\tt astro-ph/0311231}}].

\bibitem{Springel:08}
V.~{Springel}, J.~{Wang}, M.~{Vogelsberger}, A.~{Ludlow}, A.~{Jenkins},
  A.~{Helmi}, J.~F. {Navarro}, C.~S. {Frenk}, and S.~D.~M. {White}, {\it {The
  Aquarius Project: the subhaloes of galactic haloes}},  {\em MNRAS} {\bf 391}
  (2008) 1685, [\href{http://xxx.lanl.gov/abs/0809.0898}{{\tt
  arXiv:0809.0898}}].

\bibitem{Pieri:11}
L.~{Pieri}, J.~{Lavalle}, G.~{Bertone}, and E.~{Branchini}, {\it {Implications
  of high-resolution simulations on indirect dark matter searches}},  {\em
  Phys.~Rev.~D} {\bf 83} (2011) 023518,
  [\href{http://xxx.lanl.gov/abs/0908.0195}{{\tt arXiv:0908.0195}}].

\bibitem{Navarro:97}
J.~F. {Navarro}, C.~S. {Frenk}, and S.~D.~M. {White}, {\it {A Universal Density
  Profile from Hierarchical Clustering}},  {\em ApJ} {\bf 490} (1997)
  493, [\href{http://xxx.lanl.gov/abs/astro-ph/9611107}{{\tt astro-ph/9611107}}].

\bibitem{Catena:10}
R.~{Catena} and P.~{Ullio}, {\it {A novel determination of the local dark
  matter density}},  {\em J. Cosmology Astropart. Phys.} {\bf 8} (2010)
  4, [\href{http://xxx.lanl.gov/abs/0907.0018}{{\tt arXiv:0907.0018}}].

\bibitem{Salucci:10}
P.~{Salucci}, F.~{Nesti}, G.~{Gentile}, and C.~{Frigerio Martins}, {\it {The
  dark matter density at the Sun's location}},  {\em Astron.~\&~Astrophys.}
  {\bf 523} (2010) A83, [\href{http://xxx.lanl.gov/abs/1003.3101}{{\tt
  arXiv:1003.3101}}].

\bibitem{Beacom:05}
J.~F. {Beacom}, N.~F. {Bell}, and G.~{Bertone}, {\it {Gamma-Ray Constraint on
  Galactic Positron Production by MeV Dark Matter}},  {\em Physical Review
  Letters} {\bf 94} (2005) 171301,
  [\href{http://xxx.lanl.gov/abs/astro-ph/0409403}{{\tt astro-ph/0409403}}].

\bibitem{Bergstrom:05}
L.~{Bergstr{\"o}m}, T.~{Bringmann}, M.~{Eriksson}, and M.~{Gustafsson}, {\it
  {Gamma Rays from Kaluza-Klein Dark Matter}},  {\em Physical Review Letters}
  {\bf 94} (2005) 131301, [\href{http://xxx.lanl.gov/abs/astro-ph/0410359}{{\tt
  astro-ph/0410359}}].

\bibitem{Bergstrom:05a}
L.~{Bergstr{\"o}m}, T.~{Bringmann}, M.~{Eriksson}, and M.~{Gustafsson}, {\it
  {Gamma Rays from Heavy Neutralino Dark Matter}},  {\em Physical Review
  Letters} {\bf 95} (2005) 241301,
  [\href{http://xxx.lanl.gov/abs/hep-ph/0507229}{{\tt hep-ph/0507229}}].

\bibitem{Ferrer:2006hy}
  F.~Ferrer, L.~M.~Krauss and S.~Profumo,
  {\it {Indirect detection of light neutralino dark matter in the NMSSM}},
  Phys.\ Rev.\ D {\bf 74} (2006) 115007,
  [\href{http://xxx.lanl.gov/abs/hep-ph/hep-ph/0609257}{{\tt hep-ph/0609257}}].
  
\bibitem{Gustafsson:07}
M.~{Gustafsson}, E.~{Lundstr{\"o}m}, L.~{Bergstr{\"o}m}, and J.~{Edsj{\"o}},
  {\it {Significant Gamma Lines from Inert Higgs Dark Matter}},  {\em Physical
  Review Letters} {\bf 99} (2007) 041301,
  [\href{http://xxx.lanl.gov/abs/astro-ph/0703512}{{\tt astro-ph/0703512}}].

\bibitem{Goodman:11}
J.~{Goodman}, M.~{Ibe}, A.~{Rajaraman}, W.~{Shepherd}, T.~M.~P. {Tait}, and
  H.-B. {Yu}, {\it {Gamma ray line constraints on effective theories of dark
  matter}},  {\em Nuclear Physics B} {\bf 844} (2011) 55,
  [\href{http://xxx.lanl.gov/abs/1009.0008}{{\tt arXiv:1009.0008}}].

\bibitem{Profumo:10}
S.~{Profumo}, L.~{Ubaldi}, and C.~{Wainwright}, {\it {Singlet scalar dark
  matter: Monochromatic gamma rays and metastable vacua}},  {\em Phys.~Rev.~D}
  {\bf 82} (2010) 123514, [\href{http://xxx.lanl.gov/abs/1009.5377}{{\tt
  arXiv:1009.5377}}].

\bibitem{Jackson:10}
C.~B. {Jackson}, G.~{Servant}, G.~{Shaughnessy}, T.~M.~P. {Tait}, and
  M.~{Taoso}, {\it {Higgs in space!}},  {\em J. Cosmology Astropart. Phys.}
  {\bf 4} (2010) 4, [\href{http://xxx.lanl.gov/abs/0912.0004}{{\tt
  arXiv:0912.0004}}].

\bibitem{Chalons:2011ia}
  G.~Chalons and A.~Semenov,
  {\it {Loop-induced photon spectral lines from neutralino annihilation in the NMSSM}},
  JHEP {\bf 1112} (2011) 055, [\href{http://xxx.lanl.gov/abs/1110.2064}{{\tt
    arXiv:1110.2064}}].
  
\bibitem{Dudas:09}
E.~{Dudas}, Y.~{Mambrini}, S.~{Pokorski}, and A.~{Romagnoni}, {\it {(In)visible
  Z' and dark matter}},  {\em Journal of High Energy Physics} {\bf 8} (2009) 14, [\href{http://xxx.lanl.gov/abs/0904.1745}{{\tt arXiv:0904.1745}}].

\bibitem{Mambrini:09}
Y.~{Mambrini}, {\it {A clear Dark Matter gamma ray line generated by the
  Green-Schwarz mechanism}},  {\em J. Cosmology Astropart. Phys.} {\bf 12}
  (2009) 5, [\href{http://xxx.lanl.gov/abs/0907.2918}{{\tt
  arXiv:0907.2918}}].

\bibitem{Aprile:11}
E.~{Aprile}, K.~{Arisaka}, F.~{Arneodo}, A.~{Askin}, L.~{Baudis}, A.~{Behrens},
  K.~{Bokeloh}, E.~{Brown}, T.~{Bruch}, G.~{Bruno}, and et~al., {\it {Dark Matter Results from
  100 Live Days of XENON100 Data}},  {\em Physical Review Letters} {\bf 107}
  (2011) 131302, [\href{http://xxx.lanl.gov/abs/1104.2549}{{\tt
  arXiv:1104.2549}}].


\bibitem{Tempel:2012ey}
  E.~Tempel, A.~Hektor and M.~Raidal,
  \emph{Fermi 130 GeV gamma-ray excess and dark matter annihilation in sub-haloes and in the Galactic centre},
  JCAP {\bf 09} (2012) 032
  [\href{http://arxiv.org/abs/1205.1045}{arXiv:1205.1045}].
  
\bibitem{Ackermann:2012kca}
M.~Ackermann {\it et al.},   [Fermi-LAT Collaboration],
  \emph{The Fermi Large Area Telescope On Orbit: Event Classification, Instrument Response Functions, and Calibration},
  Astrophys.\ J.\ Suppl.\  {\bf 203} (2012) 4
  [\href{http://arxiv.org/abs/1206.1896}{arXiv:1206.1896}].
    
  
\bibitem{Su:2012ft}
  M.~Su and D.~P.~Finkbeiner,
  \emph{Strong Evidence for Gamma-ray Line Emission from the Inner Galaxy},
  \href{http://arxiv.org/abs/1206.1616}{arXiv:1206.1616}.

\bibitem{Hektor:2012kc}
  A.~Hektor, M.~Raidal and E.~Tempel,
  \emph{An evidence for indirect detection of dark matter from galaxy clusters in Fermi-LAT data},
  \href{http://arxiv.org/abs/1207.4466}{arXiv:1207.4466}.
  
\bibitem{Hektor:2012ev}
  A.~Hektor, M.~Raidal and E.~Tempel,
  \emph{Fermi-LAT gamma-ray signal from Earth Limb, systematic detector effects and their implications for the 130 GeV gamma-ray excess},
  \href{http://arxiv.org/abs/1209.4548}{arXiv:1209.4548}.
  
  
\bibitem{Finkbeiner:2012ez}
  D.~P.~Finkbeiner, M.~Su and C.~Weniger,
  \emph{Is the 130 GeV Line Real? A Search for Systematics in the Fermi-LAT Data},
  \href{http://arxiv.org/abs/1209.4562}{arXiv:1209.4562}.



\end{thebibliography}

\newpage

\centerline{\Large \bf  Addendum to  }

\vspace{0.5cm}

\centerline{\Large   ``Fermi 130 GeV gamma-ray excess and  dark matter annihilation  }
\centerline{\Large     in sub-haloes and in the Galactic centre"}

\vspace{0.5cm}
\centerline{\Large   E. Tempel, A. Hektor, and M. Raidal}

\vspace{0.5cm}
\centerline{\Large\bf  JCAP 1209 (2012) 032 }

\vspace{1.5cm}

\begin{figure}[h]
    \center
    \includegraphics[width=0.495\textwidth]{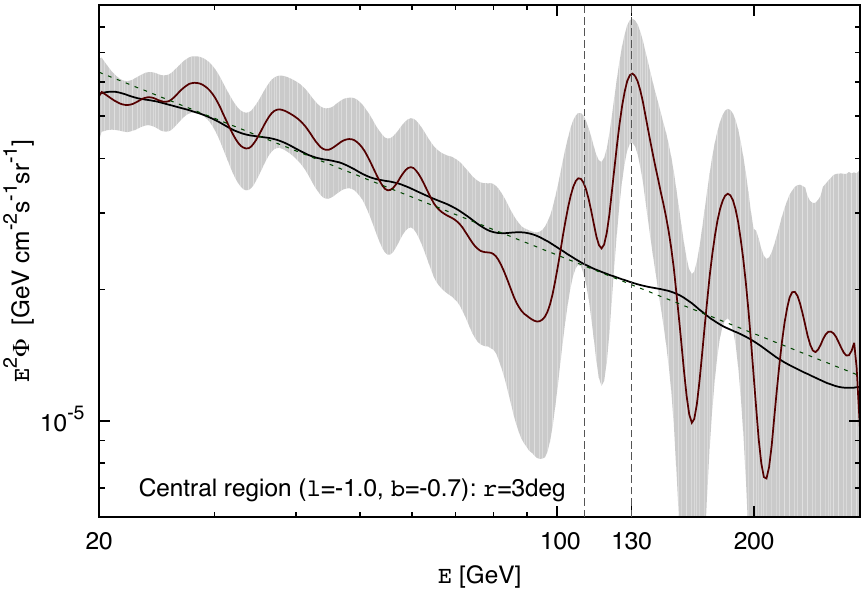}
    \caption{Updated fit to Galactic centre data using 218 week data and improved Fermi-LAT energy resolution~\cite{Ackermann:2012kca}. Vertical lines mark the 110~GeV and 130~GeV energies. }
    \label{fig_new}
\end{figure}

We have updated the fits to Fermi-LAT publicly available gamma-ray data from the Galactic centre
 presented in Ref.~\cite{Tempel:2012ey} using 218 weeks data and new improved Fermi-LAT energy resolution~\cite{Ackermann:2012kca}.
The new result is presented in Fig.~\ref{fig_new} that shows a fit to the gamma-ray flux as a function of energy together with 
$2\sigma$ error band. Compared to Figs.~3. and 4. of Ref.~\cite{Tempel:2012ey}, the most important new feature is the
presence of a double peak in the 130~GeV excess. While the previous 
Fermi-LAT energy resolution did not allow us to see the fine structure of the excess, the improved one clearly indicates
for a double peak structure, confirming similar claims made in Ref.~\cite{Su:2012ft}.

The double peak at the same energies, 110~GeV and 130~GeV, is also observed in the gamma-ray excess from 
nearby galaxy clusters~\cite{Hektor:2012kc}, suggesting that the two signals originate from the same source. 
The presence of double peak is a generic prediction of Dark Matter annihilation pattern in gauge theories, corresponding to
$\gamma\gamma$ and $\gamma Z$ final states. Thus the two seemingly unrelated gamma-ray spectra,
from the Galactic centre and from the galaxy clusters, favour the particle physics origin of the excess over any 
astrophysics origin. 

We finally note that the double peak is not present in the gamma-ray spectrum from Earth Limb~\cite{Hektor:2012ev,Finkbeiner:2012ez}.

\end{document}